\documentclass{biometrika}

\usepackage{amsmath}

\usepackage{newtxtext}
\usepackage{bbm}
\usepackage[subscriptcorrection]{newtxmath}

\graphicspath{{./art/}}

\usepackage[plain,noend]{algorithm2e}

\makeatletter
\renewcommand{\algocf@captiontext}[2]{#1\algocf@typo. \AlCapFnt{}#2} 
\def\@algocf@capt@plain{top}
\renewcommand{\algocf@makecaption}[2]{%
  \addtolength{\hsize}{\algomargin}%
  \sbox\@tempboxa{\algocf@captiontext{#1}{#2}}%
  \ifdim\wd\@tempboxa >\hsize
    \hskip .5\algomargin%
    \parbox[t]{\hsize}{\algocf@captiontext{#1}{#2}}
  \else%
    \global\@minipagefalse%
    \hbox to\hsize{\box\@tempboxa}
  \fi%
  \addtolength{\hsize}{-\algomargin}%
}
\makeatother

\def\N{\mathbb{N}}

\def\R{\mathbb{R}}

\def\var{\text{var}}

\def\ybold{\mathbf{y}}
\def\Ybold{\mathbf{Y}}
\def\xbold{\mathbf{x}}
\def\Lcal{\mathcal{L}}

\begin{document}

\jname{Biometrika}
\jyear{2024}
\jvol{103}
\jnum{1}
\cyear{2024}
\accessdate{Advance Access publication on ...} 

\received{1 March 2026} 
\revised{1 March 2026} 


\title{Using fractional derivatives to derive marginal densities}

\author{S. Y. Li}
\affil{Department of Mathematics, Imperial College London,\\ 180 Queen's Gate, London SW7 2AZ, U.K.
\email{jl1023@ic.ac.uk}}

\author{D. A. van Dyk}
\affil{Department of Mathematics, Imperial College London,\\ 180 Queen's Gate, London SW7 2AZ, U.K. \email{dvandyk@imperial.ac.uk}}

\author{\and M. Autenrieth}
\affil{Statistical Laboratory, University of Cambridge,\\ Wilberforce Road, Cambridge CB3 0WB, U.K.
\email{ma2216@cam.ac.uk}}

\maketitle

\begin{abstract}
This paper presents a novel method for analytical derivations of marginal densities using the fractional derivatives of moment-generating functions. Although the method requires likelihood functions to take specific forms, its assumptions are otherwise modest. It only requires that the prior moment-generating functions exist, are finite, and are continuous and differentiable at certain points. We also present the probabilistic and statistical insights behind this method.
\end{abstract}

\begin{keywords}
Fractional derivative; Moment-generating function; Marginal density; Minimal sufficient statistic; Bell polynomial; Conjugate prior.
\end{keywords}

\section{Introduction}
Marginal densities $p(y) = \int_{\Omega_\Theta} p(y \mid \theta)p(\theta){\rm d}\theta$ have broad applications in statistics and probabilistic analyses. Examples are marginal likelihoods in mixed-effect models and model evidence in Bayesian inference, which are central to parameter estimation and model selection. Various algorithms have been proposed to compute marginal densities. For Bayesian models, these include importance sampling \citep{zirkind1950technical}, bridge sampling \citep{meng1996simulating}, path sampling \citep{gelman1998simulating}, nested sampling \citep{skilling2006nested}, Chib's method \citep{chib1995marginal} and reversible-jump Markov-chain Monte Carlo \citep{green1995reversible}. For mixed-effect models, \citet{zeger1991generalized} proposes a Gibbs sampling approach to explore different values of the random effects $\Theta$, \citet{lindstrom1988newton} assumes multivariate normal $\Theta$ and uses Newton-Raphson and EM algorithms to estimate fixed effects, and \citet{breslow1993approximate, liu1993heterogeneity} explore Laplace approximations of the marginal densities as integrals.

These algorithms are based on approximation or sampling, which inevitably introduce arithmetic or Monte-Carlo uncertainty. Some Bayesian algorithms rely on (power-)posterior samples, but the simulation of such samples is seemingly unnecessary if parameter estimation is not of primary interest. We introduce MGF marginalisation methods: theoretical and computational tools based on (fractional) derivatives of moment-generating functions. Their importance is fourfold. First, they are analytic and therefore deliver exact marginal densities. Second, they sometimes simplify marginal density calculations. Third, they characterise a new class of likelihoods with special properties. Fourth, a subsequent paper will show that, for certain models, this is the only analytical method available.

\section{MGF-marginalisation methods}
\subsection{The number of renewals in a random interval}
\citet{cox1960number} considers a renewal process and calculates the distribution of the number $N_t$ of renewals in an interval of random length $T$, $p(N_T) = \int_0^\infty p(N_T \mid T=t)p(t){\rm d}t$ where $p(t){\rm d}t \simeq {\rm pr}(T \in [t, t+dt))$. This is an instance of a marginal density calculation in a stochastic process setting, where marginalisation over $T$ is analogous to marginalisation over parameter values in Bayesian inference, and, similarly, the distribution of $N_T$ conditioning on $T=t$ functions as the sampling distribution. \citet{cox1960number} derives formulas to find the marginal density $p(N_t)$ involving fractional derivatives; MGF-marginalisation methods inherit Cox's foundational idea.

\cite{cox1960number} focuses on the special case where the interval length $T$ follows a Gamma$(\alpha, \beta)$ distribution and the conditional distribution of $N_T$ given $T$ is discrete with support on the positive integers. His main result leverages the structure of the gamma density, 
\begin{align} \label{eqn: Cox formula number of renewals}
\begin{split}
    p(N_T=r \mid \alpha, \beta) & = \int_0^\infty L(t; r) \frac{\beta^\alpha}{\Gamma(\alpha)} t^{\alpha-1} e^{-\beta t} {\rm d}t
    = \frac{\beta^\alpha}{\Gamma(\alpha)} \int_0^\infty L(t; r) \left(-\frac{\partial}{\partial \beta} \right)^{\alpha-1} e^{-\beta t} {\rm d}t \\
    & = \frac{\beta^\alpha}{\Gamma(\alpha)} \left(-\frac{\partial}{\partial \beta} \right)^{\alpha-1} \int_0^\infty L(t; r) e^{-\beta t} {\rm d}t 
    = \frac{\beta^\alpha}{\Gamma(\alpha)} \left(-\frac{\partial}{\partial \beta} \right)^{\alpha-1} \{\Lcal L\}_t(\beta;r),
\end{split}
\end{align}
where $L(t; r)=p(N_T=r \mid T=t)=p_r(t)$ and $\{\Lcal L\}_t$ is the Laplace transform of $L$. For renewal processes, the Laplace transforms of $p_r(t)$ and $G(z, t)$ are well-known, see \citet[][4 and 5]{cox1960number}; however, their inversion into explicit formulae is generally possible only for special cases. The insight in \citet{cox1960number} is that when $T$ follows a gamma distribution, this inversion can be entirely avoided. By noticing the structure in the gamma density and how it relates to Laplace transforms, \citet{cox1960number} focuses on how the marginal probability-generating function $G(z)$ can be expressed directly as a derivative of the Laplace transform $\{\Lcal G\}_t(z, s)$, and a similar manipulation applies to $p_r$ given by \eqref{eqn: Cox formula number of renewals}. \citet[][11]{cox1960number} converted a potentially difficult contour integration problem to one of taking derivatives; our approach is in a similar vein: we derive the marginal density from $M_T(s)=\{\Lcal p \}_t(-s)$ for a given $L$ rather than from $\{\Lcal L\}_t$ for a given $p(t)$, because the latter is generally unknown for many applications of marginal densities, whereas $M_T(s)$ is often available.


In \eqref{eqn: Cox formula number of renewals}, the shape parameter $\alpha$ of the gamma distribution need not be an integer. \citet{cox1960number} states that for fractional orders of the derivative in \eqref{eqn: Cox formula number of renewals}, the definition is given on page 399 of \citet{courant1936differential}. In \S \ref{sec: LC fractional derivatives} we shall demonstrate that this definition is insufficient and the study of fractional derivatives has since become more sophisticated.

This paper adopts Bayesian terminology; $y\ge0$ denotes the observation; $\Theta>0$ is the parameter to be marginalised over with its realisation denoted by $\theta$; $p(y \mid \theta)$ characterises the sampling distribution with a parameter space $\Omega_\Theta=[0, \infty)$ unless otherwise specified; $L(\theta; y)=p(y \mid \theta)$ is the likelihood function; $p(\theta)$ characterises the prior distribution; $\Ybold = (Y_1, Y_2, \ldots, Y_n)$ is an independent identically-distributed sample of size $n$ with $p(y_i \mid \theta) = L(\theta; y_i)$; and $\ybold = (y_1, y_2, \ldots, y_n)$ is a realisation of $\Ybold$. We call our derivation of marginal densities ``\textit{MGF marginalisation}''. For MGF marginalisation methods, we are particularly interested in the prior MGF, $M_\Theta(t) = \int_{\Omega_\Theta} e^{t\theta} p(\theta){\rm d}\theta$.

\subsection{Two important examples} \label{subsec: Poisson and gamma examples}
Cox's insight in his formulation of \eqref{eqn: Cox formula number of renewals} was his reliance on the structure of the gamma density. There is an alternative method to compute marginal densities by leveraging density structures, but without any need for Laplace transforms of the likelihood $L(t; y)$ in $t$. For instance, if $(Y \mid \Theta=\theta) \sim \text{Poisson}(\theta)$,
\begin{equation} \label{eqn: Poisson simple demo}
        p(y)
        = \int_{\Omega_{\Theta}} p(y \mid \theta)p(\theta) {\rm d}\theta
        = \frac{1}{y!} E_\Theta (\theta^y e^{t\theta}) \bigg|_{t=-1}
        = \frac{1}{y!} E_\Theta \left(\frac{\partial^y}{\partial t^y} e^{t\theta}\right) \bigg|_{t=-1}
        = \frac{1}{y!} \frac{\partial^y}{\partial t^y} M_{\Theta}(t)\bigg|_{t=-1}.
\end{equation}
Similarly, if $(Y \mid \Theta=\theta) \sim \text{Gamma}(\alpha, \theta)$ where $\alpha$ is known,
\begin{equation} \label{eqn: gamma simple demo}
        p(y)
        = \int_{\Omega_{\Theta}} p(y \mid \theta)p(\theta) {\rm d}\theta
        = \frac{y^{\alpha - 1}}{\Gamma(\alpha)} E_\Theta(\theta^\alpha e^{t\theta}) \Bigg|_{t=-y}
        = \frac{y^{\alpha - 1}}{\Gamma(\alpha)} E_\Theta \left(\frac{\partial^\alpha}{\partial t^\alpha} e^{t\theta}\right) \Bigg|_{t=-y}
        = \frac{y^{\alpha - 1}}{\Gamma(\alpha)} M^{(\alpha)}_{\Theta}(-y),
\end{equation}
where $M^{(\alpha)}_{\Theta}(t) = (\partial/\partial t)^\alpha_{(-\infty)+}M_\Theta(t)$ is the $\alpha$-order Liouville--Caputo fractional derivative of the prior MGF. We define and discuss Liouville--Caputo fractional derivatives in \S \ref{sec: LC fractional derivatives}.
\begin{definition} \label{def: MGF-marginalisable at -infinity}
    For real-valued statistics $a, b, c: \R^n \mapsto \R_+$, assume the prior MGF exists, is continuous, and is differentiable up to $a(\ybold)$. If
    \begin{equation} \label{eqn: -inf-MGF-marginalisable definition}
        p(\ybold) = \int_{\Omega_\Theta} L(\theta; \ybold)p(\theta)d\theta = c(\ybold) \left(\frac{\partial}{\partial t} \right)^{a(\ybold)}_{(-\infty)+} M_{\Theta}(t) \Biggm|_{t=-b(\ybold)},
    \end{equation}
    then we say $L$ is $D_{(-\infty)+}$-MGF-marginalisable.
\end{definition}
We derive properties of $D_{(-\infty)+}$-MGF-marginalisable likelihood $L$ and their statistics $a$ and $b$ in \S \ref{subsec: statistical insight}. Examples include the Poisson likelihood by \eqref{eqn: Poisson simple demo} with $a(y)=y,\, b(y)=1,\, c(y)=(y!)^{-1}$ since integer-ordered derivatives are special cases of Liouville--Caputo fractional derivatives, the gamma likelihood by \eqref{eqn: gamma simple demo} with $a(y)=\alpha,\, b(y)=y,\, c(y)=y^{\alpha-1}/\Gamma(\alpha)$, the Rayleigh likelihood with $a(\ybold)=1, b(\ybold)=(\sum_{i=1}^n y_i^2)/2, c(\ybold) = \prod_{i=1}^n y_i$, the normal likelihood with known mean $\mu$ with $a(\ybold)=n/2, b(\ybold)=\sum_{i=1}^n (y_i-\mu)^2/2, c(\ybold)=(2\pi)^{-0.5}$ and the Gompertz likelihood with known scale $\beta$ with $a(\ybold) = 1, b(\ybold) = n-\sum_{i=1}^n e^{\beta y_i}, c(\ybold)=\beta e^{\beta \sum_{i=1}^n y_i}$.
\begin{remark}
    In the notation of \eqref{eqn: -inf-MGF-marginalisable definition}, the only assumptions of \eqref{eqn: Poisson simple demo} and \eqref{eqn: gamma simple demo} are that the prior MGF $M_\Theta(t)$ exists and is continuous around $-b(y)$, $M_\Theta(-b(y)) < \infty$, $M_\Theta^{(a(y))}(-b(y))$ exists, and $E_\Theta(\Theta^{\langle a(\ybold) \rangle + 1}) < \infty$, where $\langle x \rangle$ is the largest integer strictly less than $x$.
\end{remark}

\section{Liouville--Caputo fractional derivatives} \label{sec: LC fractional derivatives}
Following \cite{hilfer2008threefold}, for $-\infty \le u < x < v \le \infty$, the Riemann--Liouville fractional integral of order $\alpha > 0$ for a locally integrable function $f: [u, v] \to \R$ is defined as
\begin{equation} \label{eqn: RL fractional integral}
    (I_{u+}^{\alpha} f)(x) = \frac{1}{\Gamma(\alpha)} \int_u^x (x-y)^{\alpha - 1} f(y){\rm d}y.
\end{equation}
For a fractional derivative of order $\alpha > 0$, let $\langle \alpha\rangle + 1 \ge \alpha$ be the integer part of the order, and $\langle \alpha \rangle + 1 - \alpha = \gamma \in [0, 1)$ be the fractional part. A Liouville--Caputo fractional derivative of order $\alpha$ is defined to be the Riemann--Liouville integral of order $\gamma$ of the derivative of order $\langle \alpha \rangle + 1$. Namely, for $\gamma \in (0, 1)$, $f \in \rm{AC}^{\langle \alpha \rangle}((-\infty, -b])$ for some $b>0$,
\begin{equation} \label{eqn: LC derivative}
    (D_{u+}^{\alpha} f)(x) = \left(I_{u+}^{\gamma} \frac{\partial^{(\langle \alpha \rangle + 1)}}{\partial x^{(\langle \alpha \rangle + 1)}} f \right)(x)
    =\frac{1}{\Gamma(\gamma)} \int_u^x (x-y)^{\gamma - 1} f^{(\langle\alpha\rangle + 1)}(y){\rm d}y.
\end{equation}
For $\alpha = 0$, $(I_{u+}^{0} f)(x) = (D_{u+}^{0} f)(x) = f(x)$;
and for $\alpha \in \N_0$, $(D_{u+}^{\alpha} f)(x) = (\partial/\partial x)^\alpha f(x)$.
The insight is that one first differentiates $f$ to the smallest integer order greater than $\alpha$ and then remediates the extra fractional order. This avoids a problem with the seemingly simpler inverse fractional integral, $(I_{u+}^{-\alpha} f)(x)$ for $\alpha \in \R_+$, namely, that $\Gamma(-\alpha)$ can diverge. By the Hardy--Littlewood--Sobolev Theorem \citep[][\S 2.2.1.8]{hilfer2008threefold}, for $\alpha \in (0, 1)$, the fractional integral $I^\alpha_{u+}$ defined by \eqref{eqn: RL fractional integral} is a linear operator on $L^1([u, v])$. Consequently, the fractional derivative $D^\alpha_{u+}$ defined using $I^\alpha_{u+}$ is also linear on $L^1([u, v])$. Linearity of the Liouville--Caputo fractional derivatives is fundamental for our purposes, since expectations can only be swapped with linear operators.

Fractional derivatives must be studied with care, as there are different definitions, each preserving only some properties of the ordinary derivatives, while none of the definitions simultaneously preserve all properties. One can verify that, for the Liouville--Caputo fractional derivative $D_{(-\infty)+}^\alpha$ with respect to $t$, $D_{(-\infty)+}^\alpha e^{t\theta} = \theta^\alpha e^{t\theta}$; this is why Liouville--Caputo fractional derivatives are used for gamma-likelihood MGF-marginalisation in \eqref{eqn: gamma simple demo}. 
In his derivation, \citet{cox1960number} cites \citet{courant1936differential} for the definition of fractional derivatives. According to \citet{courant1936differential}, the only working definition is equivalent to $D_{0+}^\alpha$ in \eqref{eqn: LC derivative}. This specific definition is inappropriate, not only for Cox's purpose but also for the general method of finding marginal densities via Laplace transforms. Its inappropriateness is demonstrated by $D_{0+}^\alpha e^{t\theta} = \theta^\alpha e^{t\theta} \Gamma_i(\gamma, \theta t)/\Gamma(\gamma)$, which is not proportional to the gamma density of the random interval $T$, where $\Gamma_i$ is the lower incomplete gamma function.


\section{Mathematical insight}
\subsection{Statistical insight} \label{subsec: statistical insight}
We connect MGF-marginalisation to certain standard probabilistic and statistical concepts and, by identifying a general structure for $D_{(-\infty)+}$-MGF-marginalisable models, introduce a practical criterion for membership in this class. We consider the sample size $n=1$, but larger samples follow the same pattern.
\begin{theorem} \label{theorem: gamma-prior conjugate family}
    Suppose $L(\theta; y)$ is a likelihood that is continuous in $\theta$ and that the marginal density $p(y)$ exists. Then the gamma distribution is the conjugate prior of $L$ if and only if $L$ is $D_{(-\infty)+}$-MGF-marginalisable in terms of the real-valued statistics $a$, $b$ and $c$
    ; equivalently,
    \begin{equation} \label{eqn: gamma likelihood formulation}
        L(\theta; y) = c(y) \theta^{a(y)} e^{-b(y)\theta} \mathbbm{1}[\theta > 0].
    \end{equation}
\end{theorem}
\begin{proof}
    Suppose $L$ is $D_{(-\infty)+}$-MGF-marginalisable. Then 
    \begin{equation*}
        p(y) = c(y) \left(\frac{\partial}{\partial t} \right)^{a(y)}_{(-\infty)+} M_\Theta(t) \Biggm|_{t=-b(y)}
        = \int_{\Omega_\Theta} c(y) \theta^{a(y)} e^{-b(y)\theta} p(\theta) {\rm d}\theta
    \end{equation*}
    holds for any $p(\theta)$ for which $M_\Theta(t)$ exists. Hence for any bounded interval $\Omega \subset \R_+$ of length $|\Omega|$, we may consider the uniform density $p(\theta) = \mathbbm{1}[\theta \in \Omega]/|\Omega|$ and
    \begin{equation} \label{eqn: equating coefficient for uniform}
        \int_\Omega \frac{1}{|\Omega|} L(\theta; y) {\rm d}\theta
        = p(y)
        = \int_\Omega \frac{1}{|\Omega|} c(y) \theta^{a(y)} e^{-b(y)\theta} {\rm d}\theta
        \implies L(\theta; y) = c(y) \theta^{a(y)} e^{-b(y)\theta} \mathbbm{1}[\theta > 0]
    \end{equation}
    by the fundamental theorem of calculus. The converse proof is trivial by the steps outlined in \eqref{eqn: gamma simple demo}.

    We must show $L$ takes the form in \eqref{eqn: equating coefficient for uniform} if and only if the gamma distribution is its conjugate prior. If $L$ takes the form in \eqref{eqn: equating coefficient for uniform}, with a Gamma$(\alpha_0, \beta_0)$ prior distribution, the posterior density is
    \begin{equation*}
        p(\theta \mid y) \propto \theta^{\alpha_0 + a(y) - 1}e^{-\{\beta_0 + b(y)\}\theta},
    \end{equation*}
    a Gamma$(\alpha_0 + a(y), \beta_0 + b(y))$ density. Conversely, if Gamma$(\alpha_0, \beta_0)$ is the conjugate prior to an arbitrary likelihood $L^*(\theta; y)$, then the posterior is also a gamma distribution, say Gamma$(\alpha_1, \beta_1)$. But $\theta^{\alpha_1-1}e^{-\beta_1\theta} 
        \propto L^*(\theta; y) \theta^{\alpha_0-1} e^{-\beta_0\theta}
        \implies$
    \begin{equation} \label{eqn: likelihood formulation in gamma conjugacy}
        L^*(\theta; y) 
        \propto \frac{\theta^{\alpha_1-1}e^{-\beta_1\theta}}{\theta^{\alpha_0-1} e^{-\beta_0\theta}} = \theta^{\alpha_1 - \alpha_0} e^{-(\beta_1 - \beta_0)\theta}.
    \end{equation}
    The posterior parameters $\alpha_1$ and $\beta_1$ depend only on the corresponding prior parameter and likelihood, so $\alpha_1=\alpha_1(y, \alpha_0)$ and $\beta_1=\beta_1(y, \beta_0)$. The left side $L^*(\theta; y)$ depends on the parameter $\theta$ and data $y$, but does not depend on the hyperparameters $\alpha_0$, $\beta_0$; however, the right side of \eqref{eqn: likelihood formulation in gamma conjugacy} depends on $\alpha_0$ and $\beta_0$. Since proportionality must hold for any $\alpha_0, \beta_0 > 0$, the only possibility is
    $a = \alpha_1(y, \alpha_0) - \alpha_0$ does not depend on $\alpha_0$, and $b = \beta_1(y, \beta_0) - \beta_0$ does not depend on $\beta_0$, so $a$ and $b$ depend only on $y$ and $a^*(y)=a$ and $b^*(y)=b$ are statistics. Consequently, \eqref{eqn: likelihood formulation in gamma conjugacy} can be re-expressed in the form of \eqref{eqn: equating coefficient for uniform} as $L^*(\theta; y) \propto \theta^{a^*(y)} e^{-b^*(y)\theta}$ as required. 
\end{proof}
Because \eqref{eqn: gamma likelihood formulation} has the form of a gamma likelihood, any result that holds for general gamma likelihoods can be generalised to all $D_{(-\infty)+}$-MGF-marginalisable likelihoods. Transformations $\theta = \theta(\cdot)$ can be used to identify specific likelihoods with \eqref{eqn: gamma likelihood formulation}; for example, we consider the precision parameter $\theta = \sigma^{-2}$ for normal likelihoods with known mean parameter.

\begin{remark}
    The Weibull likelihood with known $\rho$, $L(\lambda; \ybold \mid \rho) = \prod_{i=1}^n\lambda \rho y_i^{\rho-1} e^{-\lambda y_i^\rho}$, for an independent sample of size $n$ is $D_{(-\infty)+}$-MGF-marginalisable $(a(\ybold)=n, b(\ybold)=\sum_{i=1}^n y_i^\rho, c(\ybold)=\rho^n \prod_{i=1}^n y_i^{\rho-1})$. The calculation is easier with a different formulation via the chain rule:
    \begin{align} \label{eqn: Weibull likelihood marginalisation}
    \begin{split}
        p(\ybold \mid \rho) 
    = & E_{\lambda}\left\{\lambda^n \rho^n \left(\prod_{i=1}^n y_i^{\rho-1}\right) e^{-\lambda \sum_{i=1}^n y_i^\rho}\right\}
    = (-1)^n E_{\lambda}\left\{\frac{\partial^n}{\partial y_1 \partial y_2 \cdots \partial y_n} e^{-\lambda \sum_{i=1}^n y_i^\rho} \right\} \\
    = & (-1)^n \frac{\partial^n}{\partial y_1 \partial y_2 \cdots \partial y_n} M_{\lambda}\left(-\sum_{i=1}^n y_i^\rho\right).
    \end{split}
    \end{align}
\end{remark}
The order of operators in \eqref{eqn: -inf-MGF-marginalisable definition} and \eqref{eqn: Weibull likelihood marginalisation} differ. The procedure in \eqref{eqn: -inf-MGF-marginalisable definition} is to compute the fractional derivatives and then substitute $t=-b(\ybold)$; \eqref{eqn: Weibull likelihood marginalisation} reverses this order by making the substitution first. By absorbing the normalising constant $c(\ybold)$ into the derivative of the MGF, \eqref{eqn: Weibull likelihood marginalisation} achieves greater simplicity. In general, this approach requires the differential equation $c(y_i)=B_m\{-b^{(1)}(y_i), -b^{(2)}(y_i), \cdots, -b^{(m)}(y_i)\}$ to hold for some $m\in\N$ for all $y_i$, where $b^{(j)}(y_i)=(\partial/\partial y_i)^j b(y_i)$ for all $j\in\{1, 2, \ldots, m \}$ and $B_m$ denotes the $m$-th complete Bell polynomial. See Appendix for the definition of Bell polynomials.

\begin{corollary} \label{corollary: sufficient statistics a,b}
    For $D_{(-\infty)+}$-MGF-marginalisable likelihoods, the statistic $(a(\ybold), b(\ybold))$ is jointly sufficient for $\theta$. If one of $a(\ybold)$ and $b(\ybold)$ is a known constant, then the other one is minimal sufficient for $\theta$. Otherwise, $(a(\ybold), b(\ybold))$ is minimal sufficient for $\theta$.
\end{corollary}
\begin{proof}
    Suppose $L$ is $D_{(-\infty)+}$-MGF-marginalisable. Consider two realisations of $\Ybold$, denoted by $\xbold$ and $\ybold$. Then by Theorem \ref{theorem: gamma-prior conjugate family}, the ratio of densities is
    \begin{equation*}
        \frac{p(\ybold \mid \theta)}{p(\xbold \mid \theta)}
        = \frac{L(\theta; \ybold)}{L(\theta; \xbold)}
        = \frac{c(\ybold) \theta^{a(\ybold)} e^{-b(\ybold)\theta} }{c(\xbold) \theta^{a(\xbold)} e^{-b(\xbold)\theta}} 
        = \frac{c(\ybold)}{c(\xbold)} \theta^{a(\ybold)-a(\xbold)} e^{-\{b(\ybold)-b(\xbold)\}\theta}.
    \end{equation*}
    This ratio is a constant function of $\theta$ if and only if $a(\xbold)=a(\ybold)$ and $b(\xbold)=b(\ybold)$. Therefore, by Theorem 6.3 in \cite{lehmann1950completeness}, $(a(\ybold), b(\ybold))$ is a minimum sufficient statistic for $\theta$. Clearly if either $a(\ybold)$ or $b(\ybold)$ is constant, the other is minimal sufficient.
\end{proof}
Corollary \ref{corollary: sufficient statistics a,b} highlights the roles of the derivative order and the point of evaluation in $D_{(-\infty)+}$-MGF-marginalisation methods. Essentially, despite the normalisation constant $c(\ybold)$, the high-order MGF derivative is an operation on the prior MGF based on the minimal sufficient statistic of the parameter being marginalised over. Corollary \ref{corollary: sufficient statistics a,b} shows how the marginal density is represented as a sufficient statistic by MGF-marginalisation methods. It also highlights that \eqref{eqn: -inf-MGF-marginalisable definition} is an alternative representation of the marginal density $p(\ybold)$ as a statistic, despite its defining integral.

\subsection{Probabilistic insight}
We consider a sample of size $n=1$ for probabilistic insight. Under the Poisson likelihood defined in Equation \eqref{eqn: Poisson simple demo}, the marginal distribution of $Y$ is discrete; we can derive its \textit{marginal probability-generating function}, namely $G_Y(s)=\sum_{y=0}^\infty s^y \int_{\Omega_\Theta} p(y\mid \theta)p(\theta)d\theta$ using the law of iterated expectation,
\begin{equation} \label{eqn: TOWER property for marginal pgf}
    G_Y(s) = E_Y\left(s^Y\right) = E_\Theta\left\{E_Y\left(s^Y \mid \Theta\right)\right\} = E_\Theta\left\{e^{-\Theta(1-s)}\right\} = M_{\Theta}(s-1)
\end{equation}
for any $s \in [0, 1)$. Therefore, the prior MGF is equivalent to $G_Y(s)$ if $(Y \mid \Theta=\theta)$ has a Poisson distribution. Differentiating the prior MGF to obtain $p(y)$ is validated by \eqref{eqn: TOWER property for marginal pgf}:
\begin{equation} \label{eqn: poisson prior mgf expansion = marginal pgf expansion}
    \frac{1}{y!}\left(\frac{\partial}{\partial t} \right)^y M_{\Theta}(t)\Bigg|_{t=-1}
    = \frac{1}{y!}\left(\frac{\partial}{\partial t} \right)^y G_Y(t+1)\Bigg|_{t=-1}
    = \frac{G_Y^{(y)}(0)}{y!}
    = p(y).
\end{equation}
The discrete nature of the Poisson distribution allows us to express \eqref{eqn: poisson prior mgf expansion = marginal pgf expansion} in terms of $G_Y$. For general $D_{(-\infty)+}$-MGF-marginalisable likelihoods, we instead consider the marginal Laplace transform.
\begin{lemma} \label{lemma: gamma likelihood marginal Laplacian insight}
    Under gamma likelihoods, $D_{(-\infty)+}$-MGF-marginalisation asymptotically coincides with the inversion formula of the marginal Laplace transform due to \cite{post1930generalized}.
\end{lemma}
\begin{proof}
    Suppose $(Y_\alpha \mid \Theta_\alpha=\theta_\alpha) \sim \text{Gamma}(\alpha, \theta_\alpha)$. Let $\Lambda=E_Y(Y_\alpha \mid \Theta_\alpha) = \alpha/\Theta_\alpha$ be arbitrarily fixed, so $\lambda = \alpha/\theta_\alpha$ and $\var_Y(Y_\alpha \mid \Theta_\alpha = \theta_\alpha) = \alpha \theta_\alpha^{-2} = \lambda^2/\alpha$. For any $\epsilon > 0$, by Chebyshev's inequality, ${\rm pr}(|Y_\alpha-\lambda| \ge \epsilon \mid \Theta_\alpha=\theta_\alpha) \le \var_Y(Y_\alpha \mid \Theta_\alpha = \theta_\alpha)/\epsilon^2 = \lambda^2/(\alpha\epsilon^2) \to 0$ as $\alpha \to \infty$. Hence $\lim_{\alpha\uparrow\infty}(Y_\alpha \mid \Theta_\alpha = \theta_\alpha) = \lambda$ in conditional probability. Suppose, given $\Theta_\alpha = \theta_\alpha$, $X_\alpha = Y_\alpha^{-1}$, so $Y_\alpha > 0$ almost surely. Since the transformation $g(y) = y^{-1}$ is continuous for all $y > 0$, by the continuous mapping theorem, $\lim_{\alpha\uparrow\infty}(X_\alpha \mid \Theta_\alpha = \theta_\alpha) = g(\lambda) = \lambda^{-1}$ in conditional probability. For any fixed $t \ge 0$, the function $h(x) = e^{-tx}$ is continuous and bounded for all $x \in \R_+$ such that $h(x)\in[0, 1]$. By the continuous mapping theorem, $\lim_{\alpha\uparrow\infty} (e^{-tX_\alpha} \mid \Theta_\alpha = \theta_\alpha) = h(\lambda^{-1}) = e^{-t/\lambda}$ in conditional probability. Now for a sufficiently large integer $\alpha$, let $X=X_\alpha, \, Y=Y_\alpha$ and $\Theta = \Theta_\alpha$. By the bounded convergence theorem with dominating constant equal to 1, for the Laplace transform $\Lcal_X(t) = \int_{\Omega_X}e^{-tx}p(x){\rm d}x$,
    \begin{align*}
        & \lim_{\alpha \uparrow \infty} E_X\left(e^{-tX_\alpha} \mid \Theta_\alpha =\theta_\alpha \right)
        = E_X\left(e^{-\frac{t}{\lambda}} \mid \Theta_\alpha = \theta_\alpha \right) 
        = e^{-\frac{t}{\lambda}} \implies \\
        & \lim_{\alpha \uparrow \infty} \Lcal_X(t) 
        = \lim_{\alpha \uparrow \infty} E_X\left(e^{-tX_\alpha}\right)
        = \lim_{\alpha \uparrow \infty} E_\Theta\left\{E_X\left(e^{-tX_\alpha} \mid \Theta_\alpha\right)\right\}
        = E_\Theta\left(e^{-\frac{t}{\Lambda}}\right)
        = E_\Theta\left(e^{-\frac{t\Theta}{\alpha}}\right)
        = \Lcal_\Theta\left(\frac{t}{\alpha} \right)
    \end{align*}
    by the dominated convergence theorem applied on the random variable $E_X(e^{-tX_\alpha} \mid \Theta_\alpha)$ with an upper bound of $1$. So for $z=t/\alpha$, $\Lcal_X(\alpha z) \sim \Lcal_\Theta(z)$ as $\alpha \to \infty$. Hence, by the inverse Laplace transform formula of \cite{post1930generalized}, the marginal density of $X$, $p_X(x)$, is
    \begin{align*}
        p_X(x)
        = & \lim_{\alpha \uparrow \infty} \frac{1}{\Gamma(\alpha)} \left(\frac{\alpha}{x} \right)^{\alpha+1} \left(-\frac{\partial}{\partial z} \right)^\alpha \Lcal_X(z) \Biggm|_{z=\frac{\alpha}{x}}
        = \lim_{\alpha \uparrow \infty} \frac{1}{\Gamma(\alpha)} \left(\frac{1}{x} \right)^{\alpha+1} \left(-\frac{\partial}{\partial z} \right)^\alpha \Lcal_X(\alpha z) \Biggm|_{z=\frac{1}{x}} \\
        = & \lim_{\alpha \uparrow \infty} \frac{1}{\Gamma(\alpha)} \left(\frac{1}{x} \right)^{\alpha+1} \left(-\frac{\partial}{\partial z} \right)^\alpha \Lcal_\Theta(z) \Biggm|_{z=\frac{1}{x}}
        = \lim_{\alpha \uparrow \infty} \frac{1}{\Gamma(\alpha)} \left(\frac{1}{x} \right)^{\alpha+1} \left(\frac{\partial}{\partial z} \right)^\alpha M_\Theta(z) \Biggm|_{z=-\frac{1}{x}}.
    \end{align*}
    Given $\Theta_\alpha = \theta_\alpha$, $X_\alpha$ is inverse gamma distributed with parameters $\alpha$ and $\theta_\alpha$, so the marginal density of $Y$, $p_Y(y)$, is
    \begin{equation*}
        p_Y(y) = p_X\left(y^{-1}\right)\Bigg|\frac{\partial y}{\partial x} \Bigg|
        = \Big|y^{-2}\Big| \lim_{\alpha \uparrow \infty} \frac{y^{\alpha+1}}{\Gamma(\alpha)} \left(\frac{\partial}{\partial z} \right)^\alpha M_\Theta(z) \Biggm|_{z=-y}
        = \lim_{\alpha \uparrow \infty} \frac{y^{\alpha-1}}{\Gamma(\alpha)} \left(\frac{\partial}{\partial z} \right)^\alpha M_\Theta(z) \Biggm|_{z=-y},
    \end{equation*}
    the $D_{(-\infty)+}$-MGF-marginalisation formula for gamma likelihoods, where $a(y)=\alpha, \, b(y)=y, \, c(y)=y^{\alpha-1}/\Gamma(\alpha)$
    \end{proof}
By Theorem \ref{theorem: gamma-prior conjugate family}, we simply need to set $a(\Tilde{y}) = \alpha$ and $b(\Tilde{y}) = y$ for Lemma \ref{lemma: gamma likelihood marginal Laplacian insight} to be generalised to $\Tilde{Y}$ following any other $D_{(-\infty)+}$-MGF-marginalisable likelihoods.

\section{Marginalisation with other lower limits}
So far we have discussed MGF marginalisation using $D_{(-\infty)+}^\alpha$, but this is not the only possibility. Considering different lower limits $u \in \R$ in \eqref{eqn: LC derivative} can yield marginalisation formulas for different likelihood forms. For example, for $u=0$, the fractional derivative with respect to $y$ is $D_{0+}^{w+1} y^{v+w+1} = \Gamma(v+w+2)/\Gamma(v+1) y^v$. This motivates a variant of Definition \ref{def: MGF-marginalisable at -infinity} with the same assumptions. If $\phi$ is a known parameter and
\begin{equation} \label{eqn: 0-MGF-marginalisable definition}
    p(\ybold \mid \phi) = c(\ybold) \left(\frac{\partial}{\partial t} \right)^{\phi}_{0+} t^{a(\ybold)} M_{\Theta}(\log(t)) \Biggm|_{t=b(\ybold)},
\end{equation}
we say $L$ is $D_{0+}$-MGF-marginalisable. 

There are fewer named likelihoods that are $D_{0+}$-MGF-marginalisable than $D_{(-\infty)+}$-MGF-marginalisable. For beta likelihoods with one known shape parameter $\alpha$, \eqref{eqn: 0-MGF-marginalisable definition} gives marginal densities integrating over the other shape parameter $\theta=\beta$ by setting $\phi = \alpha$, $b(\ybold) = \prod_{i=1}^n (1-y_i)$ and $c(\ybold) = \prod_{i=1}^n y_i^{\alpha-1}/\Gamma(\alpha)$. Similar results hold when the beta shape parameters are interchanged and for beta prime and Dirichlet likelihoods. 

By the proof of Theorem \ref{theorem: gamma-prior conjugate family} with the same steps, for sample size $n=1$, we can show that all $D_{0+}$-MGF-marginalisable likelihoods must take the form
\begin{equation} \label{eqn: beta likelihood formulation}
    L(\theta; y \mid \phi) = c(y)\frac{\Gamma(\theta+\phi)}{\Gamma(\theta)}b(y)^{\theta-1}.
\end{equation}
Unlike Theorem \ref{theorem: gamma-prior conjugate family}, the conjugate prior for \eqref{eqn: beta likelihood formulation} is not a named distribution. For independent identically-distributed samples $\xbold$ and $\ybold$ following $p(\ybold \mid \theta, \phi)$,
\begin{equation*}
    \frac{p(\ybold \mid \theta, \phi)}{p(\xbold \mid \theta, \phi)}
    = \frac{L(\theta; \ybold \mid \phi)}{L(\theta; \xbold \mid \phi)}
    = \frac{c(\ybold)\Gamma(\theta + \phi)b(\ybold)^{\theta-1}/\Gamma(\theta)}{c(\xbold)\Gamma(\theta + \phi)b(\xbold)^{\theta-1}/\Gamma(\theta)}
    =\frac{c(\ybold)}{c(\xbold)}\left\{\frac{b(\ybold)}{b(\xbold)} \right\}^{\theta-1},
\end{equation*}
which is constant with respect to $\theta$ if and only if $b(\xbold) = b(\ybold)$, showing that $b(\ybold)$ is the minimal sufficient statistic for $\theta$.


\section*{Acknowledgement}
The authors thank Prof. Xiao-Li Meng and Prof. Simon C. Harris for their encouragement to search for deeper mathematical insights and their astute probabilistic suggestions. The authors also thank Prof. Heather Battey for pointing out the important relevant work in \cite{cox1960number}.


\appendix

\appendixone
\section*{Appendix 1}
\subsection*{Bell polynomials} \label{appendix subsec: bell polynomials}
The $n$-th complete (exponential) Bell polynomial is given by \citep{bell1934exponential}:
\begin{equation} \label{eqn: Bell polynomial}
    B_n(x_1, \dots, x_n) = \sum \frac{n!}{j_1! j_2! \cdots j_n!} \left( \frac{x_1}{1!} \right)^{j_1} \left( \frac{x_2}{2!} \right)^{j_2} \cdots \left( \frac{x_n}{n!} \right)^{j_n},
\end{equation}
where the sum is over all sequences \( j_1, j_2, \dots, j_n \) of non-negative integers satisfying:
$$
j_1 + 2 j_2 + \cdots + n j_n = n.
$$
The complete Bell polynomials satisfy the following recurrence relationship \citep[][Equation 7.11b]{bell1934exponential}, which simplifies the computation of Bell polynomials: $\forall n \in \N_0$,
\begin{equation} \label{eqn: recursion in Bell polynomials}
    B_{n+1}(x_1, \dots, x_{n+1}) = \sum_{i=0}^n \binom{n}{i} B_{n-i}(x_1, \dots, x_{n-i}) \, x_{i+1},
\end{equation}
where \( B_0 := 1 \) by convention.

\bibliographystyle{biometrika}
\bibliography{paper-ref}

\end{document}